# WATUNet: A Deep Neural Network for Segmentation of Volumetric Sweep Imaging Ultrasound

Donya Khaledyan [1*], Thomas J. Marini[2], Avice O'Connell[2], Steven Meng[2], Jonah Kan[2], Galen Brennan[2], Yu Zhao[2], Timothy M. Baran[2], Kevin J. Parker[1,2]

[1] Department of Electrical and Computer Engineering
University of Rochester, Rochester, NY, USA

[2] Department of Imaging Sciences
University of Rochester Medical Center, Rochester, NY, USA

* Corresponding author: dkhaledy@ur.rochester.edu

## Abstract

*Objective.* Limited access to breast cancer diagnosis globally leads to delayed treatment. Ultrasound, an effective yet underutilized method, requires specialized training for sonographers, which hinders its widespread use. *Approach.* Volume sweep imaging (VSI) is an innovative approach that enables untrained operators to capture high-quality ultrasound images. Combined with deep learning, like convolutional neural networks (CNNs), it can potentially transform breast cancer diagnosis, enhancing accuracy, saving time and costs, and improving patient outcomes. The widely used UNet architecture, known for medical image segmentation, has limitations, such as vanishing gradients and a lack of multi-scale feature extraction and selective region attention. In this study, we present a novel segmentation model known as Wavelet_Attention_UNet (WATUNet). In this model, we incorporate wavelet gates (WGs) and attention gates (AGs) between the encoder and decoder instead of a simple connection to overcome the limitations mentioned, thereby improving model performance. *Main results.* Two datasets are utilized for the analysis. The public "Breast Ultrasound Images" (BUSI) dataset of 780 images and a VSI dataset of 3818 images. Both datasets contained segmented lesions categorized into three types: no mass, benign mass, and malignant mass. Our segmentation results show superior performance compared to other deep networks. The proposed algorithm attained a Dice coefficient of 0.94 and an F1 score of 0.94 on the VSI dataset and scored 0.93 and 0.94 on the public dataset, respectively. *Significance.* The experimental findings demonstrate that the proposed WATUNet model achieves precise segmentation of breast lesions in both standard-of-care and VSI images, surpassing state-of-the-art models. Hence, the model holds considerable promise for assisting in lesion identification, an essential step in the clinical diagnosis of breast lesions.

**Keywords:** Breast Ultrasound, UNet, VSI, WATUNet.

## 1. Introduction

Breast cancer is one of the most fatal forms of cancer and has become a significant public health concern (1). It is the second most common cause of cancer-related deaths worldwide in women (2). Early detection and treatment of breast cancer are critical to increase the chances of survival and prevent metastasis. Ultrasound imaging is a first-line diagnostic tool for detecting breast cancer. It is a safe, portable, cost-effective, and non-invasive imaging modality that uses high-frequency sound waves to create images of breast tissue. Ultrasound imaging may be the only diagnostic option available in low and middle-income countries (LMICs). However, a significant challenge to the widespread use of ultrasound is the need for trained sonographers, who require wide-ranging skills that can take months to years to acquire (3). A potential solution to this challenge is to adopt volume sweep imaging (VSI) (4, 5). VSI is a recently developed imaging technology that has the potential to revolutionize access to medical imaging for breast evaluation, especially in resource-limited settings (6, 7, 8). In addition to being affordable compared to other imaging modalities, including standard-of-care ultrasound, VSI enables inexperienced operators to acquire high-quality ultrasound images using a standardized imaging protocol, which requires minimal training (3, 7). VSI has been clinically tested for breast, obstetrics, lung, thyroid, and right upper quadrant scanning indications, showing promising results both in the United States and Peru (4, 9, 10). Unlike traditional ultrasound, which demands highly skilled sonographers to operate the equipment and interpret the images, VSI utilizes a simplified approach based on external body landmarks (4). With VSI for breast ultrasound, an operator starts by positioning the ultrasound probe on the surface of the breast over a palpable breast lump, then sweeps the probe in a uniform pattern while maintaining contact with the skin. This produces a series of video clips that cover the entire target region (the palpable breast lump). Then, these video sweeps are sent to radiologists or computer aided diagnostic systems for further interpretation.

The use of ultrasound imaging for detecting and diagnosing breast cancer has been significantly improved with the introduction of machine learning (ML) and deep learning (DL) techniques. Compared with other ML techniques, DL algorithms, such as convolutional neural networks (CNNs), show increased potential and reliability in diagnosing disease (5, 11, 12, 13). In the case of segmentation, these algorithms can be trained on large datasets to find boundaries by recognizing their features, such as shape, texture, and color. The DL models can also learn to accurately distinguish the region of interest from the background and other structures in the image. In addition to improving the accuracy of detection and diagnosis, DL algorithms can also help reduce the time and cost associated with traditional methods of diagnosis. This is particularly significant in resource-constrained settings with limited access to trained medical professionals and expensive diagnostic equipment.

The task of image segmentation involves categorizing pixels in an image with labels that describe their meaning. This can be achieved through semantic segmentation, which involves labeling individual pixels with object categories (14). Another approach is instance segmentation, which involves separating individual objects in the image (15). Panoptic segmentation combines both semantic and instance segmentation (16). Semantic segmentation is typically more difficult than whole-image classification because it requires labeling each pixel with a specific category rather than predicting a single label for the entire image.

This paper presents a novel framework for semantic segmentation of breast tissues in ultrasound images in both standard-of-care and VSI imaging. We propose a novel segmentation architecture known as WATUNet, where the combination of wavelet decomposition and attention mechanism replaces the plain connection between encoder and decoder in the UNet structure in order to overcome the issues with plain skip connection, and to extract more features from ultrasound images. The proposed framework includes two stages. First, image enhancement and preprocessing will be applied to the inputs. Next, we will utilize our proposed segmentation framework to analyze the images and segment the mass area. Integration of VSI and WATUNet would enable the potential for rapid automatic diagnosis of palpable breast lumps without a radiologist or a sonographer (5, 17).

The manuscript is divided into several sections with section 1 being this introduction. Section 2 outlines a brief review of some of the previous works to develop an understanding of the earlier and current state of knowledge in the image segmentation field. In section 3, we describe the datasets that were used in this study, along with the various preprocessing steps that were employed to ensure the integrity and reliability of the data, results of these sections will be discussed in Appendix C. In section 4, we introduce our novel WATUNet architecture, providing a detailed overview of the model's structure and operation. Additionally, we discuss its unique features and highlight its expected contributions to the field. In section 5, we present the results of our experiments and analyze these findings in depth. Finally, we conclude the study with a discussion of the implications of our results and reflect on the broader significance of this work within the context of the field.

## 2. Literature review

Image segmentation is a significant task in the fields of computer vision and image processing, and has wide-ranging applications such as medical image analysis, scene comprehension, robotic perception, video surveillance, image compression, and augmented reality. Numerous image segmentation methods have been developed over time, starting from the earliest approaches such as thresholding (18), region growing (19), and clustering (20), to more sophisticated methods such as active contours (21), graph cuts (22), and Markov random fields (23). However, in recent years, DL models have emerged as a new generation of image segmentation models with significant performance, often achieving the highest accuracy rates on commonly used benchmarks (24). This has led to a fundamental change in the field of image segmentation.

In their work, Noh et al. (25) presented a semantic segmentation method called DeConvNet, which consists of two main parts: an encoder that utilizes convolutional layers from the VGG16 layer network, and a multilayer deconvolutional network that takes in the feature vector and generates a map of class probabilities that are precise to the pixel level. One limitation of this work is that the DeConvNet model relies on the use of a pre-trained VGG16 network as the encoder, which may not be the most optimal choice for all types of images and datasets. Using a fixed pre-trained network may limit the model's ability to adapt to new types of images and features that are specific to a particular task or domain. In addition, the DeConvNet model does not incorporate any attention mechanisms or other advanced techniques for enhancing feature representation, which may limit its ability to accurately segment complex objects or scenes.

Badrinarayanan and colleagues introduced SegNet (26), which is based on encoder-decoder structure for semantic segmentation. Similar to the deconvolution network, SegNet's core trainable segmentation engine also contains an encoder network that has the same structure as the 13 convolutional layers in the VGG16 network. Additionally, it has a decoder network and a pixel-wise classification layer. SegNet's uniqueness lies in the way the decoder performs nonlinear upsampling of its lower-resolution input feature map(s). SegNet utilizes the pooling indices calculated during the maxpooling step of the corresponding encoder to perform this task. The utilization of maxpooling for pooling indices computation, while widely used in CNNs, can have both beneficial and detrimental effects on the accuracy of segmentation output. Specifically, the use of maxpooling in this context may result in loss of crucial spatial information and cause blurring or misalignment of object boundaries, thereby limiting the accuracy of segmentation (27).

One of the limitations of the encoder-decoder structure for segmentation is that the decoder might lose fine-grained spatial information during the down sampling process, which can result in lower segmentation accuracy. This limitation is addressed in the UNet architecture by introducing skip connections that directly connect the corresponding encoder and decoder layers. These connections allow the decoder to access and use the detailed spatial information from the encoder, resulting in more accurate segmentation. The skip connections provide a way to recover the lost spatial information by fusing the feature maps from the encoder and decoder. Ronneberger et al. (28) proposed the UNet for efficiently segmenting biological microscopy images. The UNet architecture includes two parts: a contracting path to capture context, and a symmetric expanding path to enable precise localization. The UNet training strategy relies on the use of data augmentation to learn effectively from very few annotated images. It was trained on 30 transmitted light microscopy images, and it won the 2015 International Symposium on Biomedical Imaging Cell Tracking Challenge. Over time, since the introduction of the UNet architecture, numerous adaptations of the original model have been developed to address various types of images and issues, for example, UNet++ (29, 30), UNet3+ (31), Weighted Res-UNet (32), Sharp UNet (33), Attention UNet (34), and Sharp Attention UNet (35).

UNet++ was designed to achieve more precise segmentation compared to the UNet model with plain skip connections by combining multiple UNets of different depths. The decoders of these UNets are interconnected using redesigned skip pathways that address two significant issues with the original UNet design: the uncertainty regarding the optimal depth of the architecture, and the overly restrictive design of the skip connections. Notwithstanding its superior performance, the complexity of the UNet++ model is relatively high, rendering it unsuitable for real-time processing applications. Furthermore, the increased complexity of the model results in a higher risk of overfitting, which can adversely affect its generalization performance on new datasets. The model also fails to capture adequate information from the entire range of scales, leaving ample room for further enhancement. To address this limitation, UNet3+ (31) was proposed. UNet3+ takes advantage of full-scale skip connections and deep supervisions to incorporate low-level details with high-level semantics from feature maps at different scales. UNet3+ is especially suited for organs that appear at varying scales. Nevertheless, when confronted with small objects within the image, this model demonstrates a decline in accuracy.

The Sharp UNet architecture (33) consists of a standard UNet encoder-decoder structure, with the addition of a sharpening module in between the encoder and decoder layers. This sharpening filter performs a convolution operation on the encoder feature maps before fusing them with the decoder features. It helps to make the encoder and decoder features semantically less dissimilar. We recently (35) introduced a model known "Sharp Attention UNet" which is an extension of the UNet architecture. The Sharp Attention UNet incorporates attention mechanisms to emphasize relevant regions and sharpening filters to enhance image details for lesion segmentation within ultrasound images. This approach is refined for VSI sweeps and is now termed WATUNet, the subject of this paper explained in the following sections.

## 3. Dataset and data preprocessing

### 3.1 Data collection

This study utilized two ultrasound breast imaging datasets for analysis. The following section provides a detailed description of each dataset.

#### 3.1.1 BUSI dataset

The first dataset used in this study is known as "Breast Ultrasound Images" (BUSI) (36). The dataset consists of 780 images from 600 female patients between the ages of 25 and 75, collected in 2018. Patients were scanned using a LOGIQ E9 ultrasound system and lesions were segmented with manually traced masks from the radiologist's evaluation. The images were classified into three groups: (1) 133 normal images without masses, (2) 437 images with benign masses, and (3) 210 images with malignant masses. The images are in PNG format, have varying heights and widths, and an average size of 600 × 500 pixels. The data was preprocessed by removing non-image text and labels.

### 3.1.2 VSI dataset

The VSI dataset was obtained from a clinical study performed at the University of Rochester (4). Patients with a palpable breast lump underwent a VSI examination using the Butterfly iQ ultrasound probe (Butterfly Network, Guilford) employing the small organ preset. A medical student was trained for less than 2 hours on the VSI protocol. The first step in the protocol is to mark the palpable area with an "X" as shown in Figure 1. The patient lies supine with their arm above their head, and the marked area is scanned with eight sweeps in transverse, sagittal, radial, and anti-radial orientations to image the mass in different planes, resulting in 8 separate attempts to acquire a diagnostic image, greatly increasing the chance of obtaining at least one diagnostic view. This protocol takes minimal time to learn, typically 1-2 hours. Patients are scanned with a breast preset, and operators do not change any probe settings from the preset. The operator does not interpret the image, and VSI is ideally performed focusing on the sweep over the target region, not the ultrasound screen. The exam is short and can be completed within 10 minutes, including setup. Full details of the VSI protocol are included in Marini et al (4).

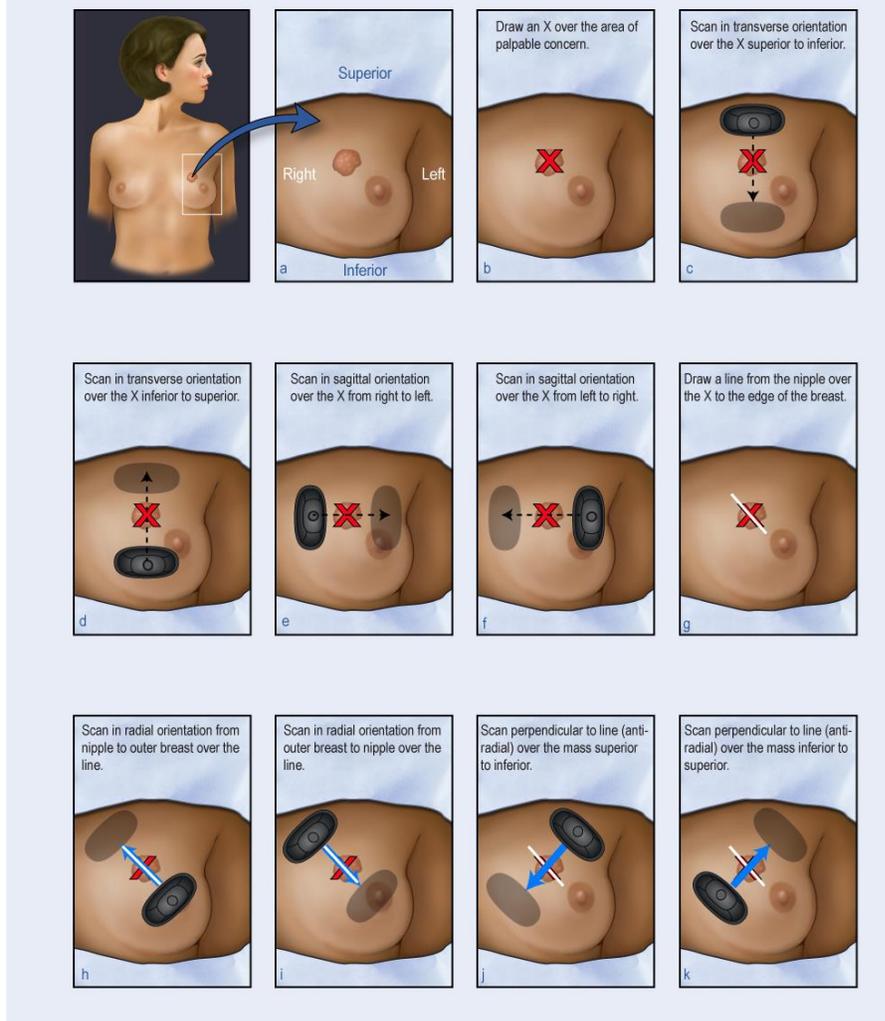

*Figure 1 Demonstration of the technique for conducting breast volume sweep imaging (VSI). Ultrasound probe sweeps are executed in various orientations, including transverse, sagittal, radial, and anti-radial. The picture is adopted from: (5)*

Each cine sweep averages 9 seconds, and the frame rate is typically 17 frames per second. The frames near the center of the sweep were selected for the segmentation as the tumors or abnormalities are more likely to be detected near the center of the ultrasound sweep. The individual frames from the center of the sweep were meticulously segmented, with every other frame being processed using the "image segmentator" application in MATLAB (version 2022b, MathWorks, Natick, MA, USA). No mass or abnormality was found in 70 cases. In 52 cases, benign masses were found, and in 20 cases malignant masses were found. Our study includes recorded biopsy data, though biopsies were not always performed.

The dataset is a three-class dataset consisting of 3818 frames with 2048 frames with no masses, 820 frames with benign findings, and 950 frames with malignant findings.

If a sonographic mass was absent, the pathology was deemed benign, and thus no biopsy was carried out. For patients with benign-appearing masses, follow-up imaging was sometimes conducted in lieu of biopsy. To classify pathology as benign in those cases, we relied on the standard of care image being assigned a BI-RADS (Breast Imaging Reporting and Data System) 2, which indicates a zero percent chance of cancer (100% chance of being benign). Pathologically, masses with a BI-RADS 3 classification hold a 98% chance of being benign, with the remaining 2% indicating potential malignancy. Therefore, although these masses were likely benign, we classified them as unknown. Notably, we did not encounter any BI-RADS 4 classifications in our study that were not biopsied. Additional information regarding pathology and its determination was previously published (4).

### 3.2 Data enhancement and augmentation

Applying image enhancement techniques to medical images, especially ultrasound images, before feeding them to DL models can help to improve the accuracy of the models. Ultrasound images can suffer from various artifacts, such as speckle noise, attenuation, and poor contrast, which can affect the quality of the image and make it difficult to interpret. Image enhancement techniques can be applied to remove or reduce these artifacts and improve the overall quality of the image. DL models can learn to identify and classify features in images based on patterns in the pixel values. By enhancing the image quality, the patterns become clearer, and the model can more accurately identify the features in the image.

Similar to our previous model (35), in this study we applied contrast limited adaptive histogram equalization (CLAHE) (37), which is an image enhancement method that improves upon the adaptive histogram equalization (AHE) (38) method by addressing the issue of excessive contrast levels. AHE can excessively enhance contrast, but CLAHE resolves this issue by setting a limit to the contrast using a histogram.

Data augmentation is a technique used to increase the size of a training dataset by modifying existing data samples. It helps to prevent overfitting by introducing more variability and diversity into the training data. This technique aims to create a more representative and generalized set of training data, particularly in the medical imaging field where data scarcity is a concern. However, it is important to carefully select augmentation parameters to avoid negative impacts on diagnostic accuracy (39). For ultrasound images, extreme brightness or zoom adjustments may lead to the loss of important details or distortions that can affect model predictions. In our study, augmentation parameters include random rotation, zoom, horizontal flip, width and height shifts, shear transformation, and brightness adjustment. These parameters define data generators that generate augmented data samples during model training. By applying these parameters randomly to the training data, the augmented dataset better represents real-world scenarios and reduces overfitting. In the results section, the performance of the segmentation model with and without data augmentation will be presented.

## 4. Methods

In this section we will explore model optimization techniques and network fine tuning. Figure 2 shows the model optimization and training workflow. This diagram illustrates the dataset preparation including masked frame extraction, data enhancement, and data augmentation. Then we trained the model while the shuffle was on, finally testing the model with unseen data.

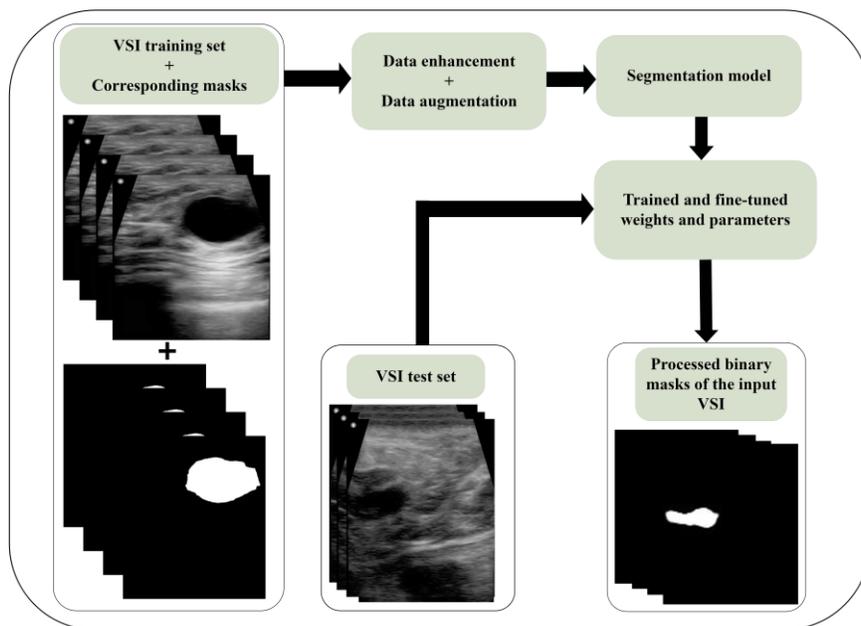

*Figure 2 Schematic depiction of the model optimization and training workflow. Workflow depicting the sequential stages of the proposed approach. In the first step, radiologists perform frame segmentation, followed by data preprocessing in the second step, encompassing data enhancement and augmentation. The preprocessed dataset is then utilized for training the proposed model. Subsequently, the model's performance is evaluated using the designated test set.*

### 4.1 Model optimization

Model optimization is an essential part of improving the performance of DL models. One significant instance is the backpropagation technique, intricately intertwined with the optimization process via gradient descent. However, backpropagation can be slow, therefore fine tuning and optimization can help this process to be faster. Another important instance is optimization's role in addressing the issues of overfitting and underfitting, which ultimately enhances the model's ability to make accurate predictions on new data. One optimization technique is exploring different activation functions for model optimization. Previously, ReLU was a widely used activation function, but considering its limitations, we tried alternatives such as Leaky ReLU, swish, and mish. Swish is a smooth function that has shown promising results in certain architectures. Mish is another activation function which is non-monotonic and has self-regularization properties. Both two activation functions exhibit negative curvature and smoothness, enabling efficient optimization and faster convergence in deep neural networks. We described each one of these activation functions in a previous work (35).

Dropout is another regularization technique in neural networks that prevents overfitting. It randomly deactivates neurons during training. This encourages the network to learn more generalized features and improve performance on new data. Furthermore, dropout reduces co-adaptation between neurons by randomly deactivating them. This will lead to preventing excessive specialization and overfitting.

The optimal values for dropout in neural networks vary depending on factors like dataset, network architecture, and training method. Gal et al. (40) found that higher dropout probabilities are generally better for deeper layers. The specific task and dataset influence the optimal dropout probability. In this study, the optimal dropout values were chosen as 0.1 for the encoder and 0.5 for the decoder. These values were based on the recommendations of Gal et al. (40) and fine-tuned through an iterative trial-and-error process.

## 4.2 Proposed network

UNet architecture is a CNN model that uses an encoder-decoder network with skip connections to preserve high-resolution information from the input images. UNet structures have been widely used for medical image segmentation, including breast tumor segmentation in ultrasound images, due to their unique architecture and outstanding performance (41). UNet-based models can be trained using relatively small datasets, which is particularly relevant in medical imaging applications where large, annotated datasets are often not readily available (41).

This skip connection in a UNet based model helps in preserving the spatial information lost during downsampling in the encoder. Concatenation of the encoder and decoder information helps preserve spatial information while increasing the depth of the feature maps, facilitating better learning of the spatial relationship between the features. In a neural network during backpropagation, the gradient is computed starting from the output layer of the network and moving backwards towards the input layer (42). The gradient values are then used to update the weights of the network using an optimization algorithm such as stochastic gradient descent or Adam. The vanishing gradient problem occurs when the gradient values become very small during backpropagation, making it difficult to update the weights in the lower layers (42). This can occur in deep networks because the gradient values are multiplied by the weight matrix at each layer, and if the weights are small, the gradient values can become exponentially small as they are propagated backwards through the network (43). As a result, the lower layers of the network may not learn effectively, and the overall performance of the network can be compromised.

Using a plain skip connection in UNet may lead to the problem of vanishing gradients during training (44). As noted earlier, this is because the gradients of the loss function with respect to the weights of the encoder and the decoder need to be propagated through the skip connections. If the plain skip connection is too deep, the gradients may become too small to be useful during training, which can slow down or even prevent convergence.

To overcome this problem, instead of a simple connection, we applied a discrete wavelet transform and attention map to highlight spatial and frequency information present in the ultrasound images. Adding a wavelet gate (WG) and an attention gate (AG) to a UNet model in place of a plain skip connection can prove beneficial. A WG can help the model capture multi-scale features in the input image. Wavelet transforms decompose the image into different frequency bands, which can capture details at different scales (45). By including a WG into the UNet model, the model can selectively attend to different frequency bands, allowing it to capture both fine-grained and coarse-grained details.

Moreover, incorporating an AG mechanism can effectively enhance the model's ability to attend to the most salient features within the input image. This is particularly useful in image segmentation tasks, where the model needs to identify the boundaries of objects accurately. By incorporating an AG into the UNet model, the model can learn to selectively attend to the most relevant regions of the input image, while down-weighting less important regions. By combining these two types of information from AGs and WGs, the UNet model can achieve better performance in image segmentation tasks. This model is designed to easily integrate into standard CNN architectures, such as the UNet model, with minimal operational overhead. Additionally, our experiment and results show that incorporating WGs and AGs in the model increases sensitivity and prediction accuracy.

### 4.2.1 Attention gate (AG)

The attention mechanism is a method to highlight high-importance features and downplay low importance features in neural network models. Oktay et al. (34) proposed a novel UNet algorithm with an attention mechanism to segment the pancreas in computed tomography (CT) images. Their proposed network consisted of three main modules: an encoder module that receives CT input images to obtain the feature map, an attention module that consists of two parallel convolutional blocks that learn to capture both local and global contextual information from the encoder path, and a decoder module that restores the concatenated feature map from the AG to the original size of the input images. The authors reported a Dice coefficient of 79.8% for their method, compared to 74.7% for UNet without attention.

Recent studies have demonstrated that the use of AGs in DL models can lead to improved network performance (46, 47). The AG architecture proposed in our study is illustrated in Figure 3. The architecture is inspired by the attention gate utilized Oktay et al. (34).

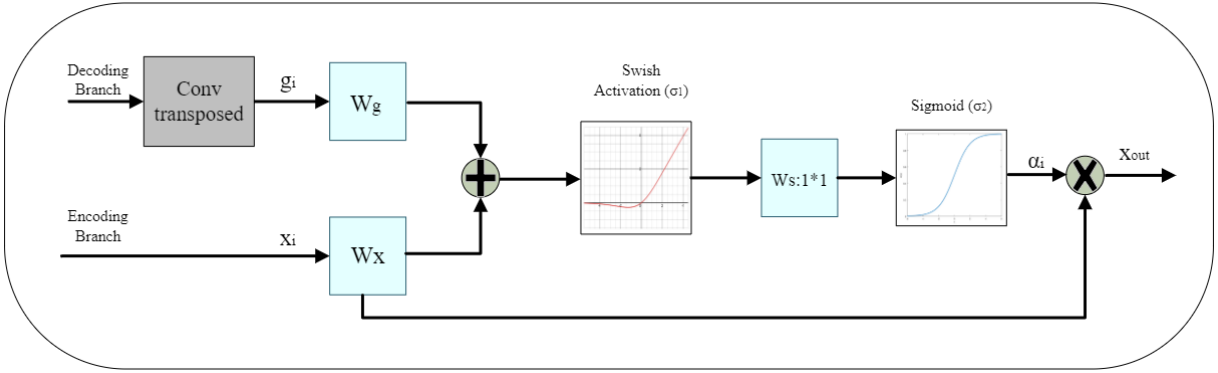

*Figure 3 Illustration of the proposed additive attention gate (AG), which employs a gating signal (g) derived from applying a transposed convolution on coarser scale features and the features from the encoding path to analyze the activations and contextual information for selecting spatial regions.*

This AG unit receives two inputs, the decoder branch, and the encoder branch. At the encoder branch, convolutional layers utilize a hierarchical approach to gradually extract high-level image features by processing local information layer by layer. This leads to a separation of pixels in a high dimensional space according to their semantics. This sequential processing of local features allows the model to condition its predictions on information collected from a large receptive field. The output of layer $l$ is the feature-map $x^l$, which captures the extracted high-level features at that layer. The subscript $i$ at $a_i, x_i$, and $g_i$ denotes the spatial dimension, where $i$ is each pixel's value.

First, we apply transposed convolution on the decoding branch, the result is $g_i$. It is used to determine focus regions while $x^l$, the feature map from the encoding path, contains local information about the image. Then, we apply a convolution of kernel size and stride of 1 with the number of channels the same as the number of feature maps of $g_i$ and $x^l$. To combine $g_i$ and $x^l$ we summed them to obtain the gating coefficient. The additive attention method may require more computational resources but has been demonstrated to achieve better accuracy compared to the multiplicative attention method (48). The resulting feature map $x_{out}$ is obtained by performing element-wise multiplication between the input feature map $x^l$ and attention coefficients $a_i$, which is expressed mathematically as:

$$x_{out} = x_i^l * a_i \qquad (1)$$

where $a_i$ can be expressed as:

$$a_i = \sigma_2(W_s^T(\sigma_1(W_x^T x_i^l + W_g^T g_i + b_g)) + b_s) \qquad (2)$$

where $\sigma_1$ is the swish function (49). Function $\sigma_2$ is the sigmoid function. The feature map $x^l$ and the gating signal vector $g_i$ are transformed linearly using a 1*1 channel-wise convolution. The parameters $W_x$, $W_g$, and $W_s$ are transformed linearly and are trainable parameters. The bias terms are $b_g$ and $b_s$. To reduce the complexity, we set the bias values to zero. Experiments have shown that setting these two values to zero does not affect the model's performance negatively.

### 4.2.2 Wavelet gate (WG)

Images include both spatial and frequency information, and wavelet transform is a valuable tool for analyzing the spatial frequency and spatial content of an image. In wavelet transform, the image is decomposed into different frequency bands or scales, each corresponding to a different level of detail. This enables the ability to analyze the image's frequency content at different scales and resolutions. In ultrasound imaging, the spatial frequency information is rich and can be leveraged to diagnose or detect patterns with higher accuracy. The continuous wavelet transform (CWT) of a signal *f(t)* is given by:

$$W_f(a,b) = \int_{-\infty}^{\infty} f(t) \frac{1}{\sqrt{a}} \psi^*(\frac{t-b}{a}) dt \qquad (3)$$

where $a$ and $b$ are the scale and translation parameters, respectively, and $\psi^*$ is the complex-valued mother wavelet function. The discrete wavelet transforms (DWT) of a signal $f[n]$ is obtained by applying a series of high-pass and low-pass filters to the signal, followed by decimation. The DWT can be represented mathematically as:

$$W_{j,k}^L = \sum_n f[n] \cdot \emptyset_{n-2^j,k} \qquad (4)$$

$$W_{j,k}^H = \sum_n f[n] \cdot \Psi_{n-2^j,k} \qquad (5)$$

where $j$ is the scale index and $k$ is the translation index. The superscripted variables $L$ and $H$ indicate whether the wavelet coefficient corresponds to the low-frequency (approximation) coefficients or high-frequency (detail) coefficients, respectively. $\emptyset$ and $\Psi$ are the scaling and wavelet functions, respectively. The scaling function is a low-pass filter that captures the coarse-scale components of a signal or image, while the wavelet function is a high-pass filter that captures the fine-scale details. Together, they provide a multi-resolution representation of the signal, in which each level of the decomposition captures a different scale of features.

The present study utilizes the DWT function to apply the wavelet transform to input images. This implementation utilizes the PyWavelets library in Python for executing the wavelet transform. PyWavelets is a free, open-source library that provides convenient tools for performing wavelet analysis in Python. The DWT function executes a four-band DWT decomposition of the image, producing four sub-bands, namely the approximation coefficients (LL) and the horizontal (LH), vertical (HL), and diagonal (HH) detail coefficients. The output of the DWT function is a tuple consisting of the LL sub-band and a concatenated tensor of the LH, HL sub-bands. The HH sub-band was considered non-essential in our dataset due to its primarily noise-related nature, and therefore its omission had no discernable effect on the results of our analyses. Figure 4 illustrates the wavelet gate proposed in our study. The LL sub-band serves as the input to the AG gate, whose architecture is shown in Figure 3. This sub-band corresponds to the low-frequency components of the original image, obtained via a low-pass filter applied to the image. Because the LL sub-band is capturing the overall structure or context of the input, this makes it a fitting input to the AG. Moreover, the LL sub-band can reduce the impact of artifacts across the network layers during the initial stages of training. The HL and LH sub-bands include high-frequency components of the original image, and therefore contain more detailed information compared to the LL sub-band. The concatenation of the AG output and these two sub-bands can help the model to better capture the important features and patterns in the data, while filtering out irrelevant or noisy information. Additionally, the AG may help to reduce the effects of dimensionality reduction that may have occurred due to the wavelet transformation.

Wavelet families include Biorthogonal, Coiflet, Harr, Symmlet, Daubechies, and others (50). There is no universally "correct" method to choose which wavelet function to use; optimality depends on the specific application (51). While the Haar wavelet is simple to compute and comprehend, the Daubechies algorithm is more complex and requires more computation. However, the Daubechies algorithm can capture details that are missed by the Haar wavelet. Thus, selecting a wavelet that closely matches the signal being processed is crucial in wavelet applications (52). For wavelet decomposition, the Daubechies wavelet with two vanishing moments provided the best results.

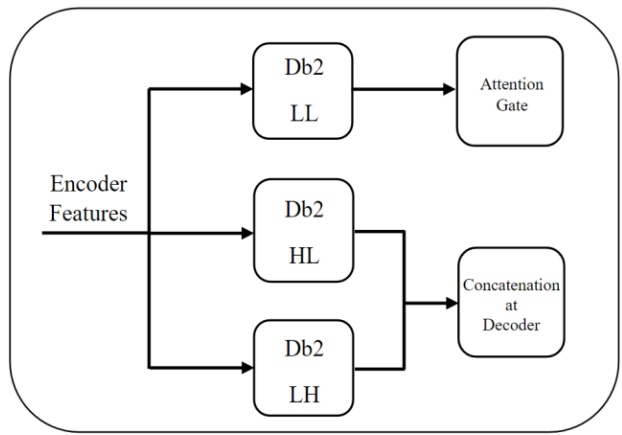

*Figure 4  Illustration of the proposed wavelet gate (WG), which employs Db2 from the PyWavelets library on the feature map from the encoder path to analyze the spatial and frequency information of the input.*

### 4.3 Network architecture

In the UNet structure, the encoder path extracts high-level features from the input image, followed by a decoder path that upsamples the feature maps to reconstruct the original image. In this paper the encoder path consists of four downsampling blocks, where each block applies convolutional layers with 3*3 kernels and a stride of 1, followed by a batch normalization layer, and a swish activation function. The downsampling is achieved by applying max pooling with a 2*2 kernel size. The output of each block is then passed through a dropout layer to prevent overfitting.

The decoder path consists of four upsampling blocks, where each block applies a transpose convolution operation with a 3*3 kernel and stride of 2, followed by concatenation with the corresponding feature maps from the AG and WG, which is applied on the encoder path features. The concatenated feature maps are then passed through a convolutional block with the same architecture as in the encoder path. Similar to the encoder path, the output of each block is then passed through a dropout layer to prevent overfitting. The final output of the model is a binary mask that represents the segmented regions of the input image. Figure 5 shows the proposed architecture.

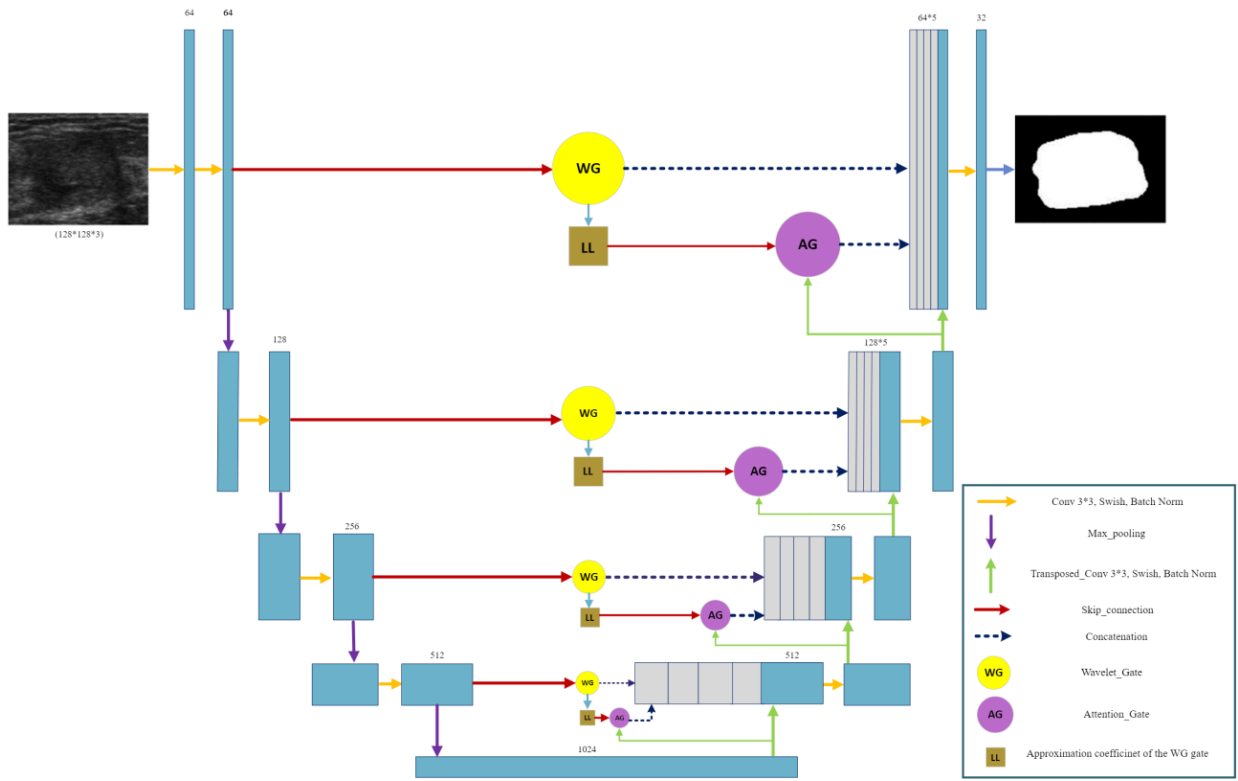

*Figure 5 The proposed WATUNet architecture with AG and WG as skip connection to provide complementary information to the UNet model.*

### 4.3.1 Loss function

In this study we utilized a custom loss function which is specifically designed for training neural networks in segmentation tasks. The custom loss function combines two commonly used loss functions: binary cross-entropy (BCE) loss and Dice loss. The BCE loss is typically used for binary classification problems such as image segmentation, where each pixel is classified as foreground or background. It quantifies the discrepancy between predicted probabilities and true binary labels. More details are explained in appendix A.

## 5. Results and Discussion

Within this section, we assessed the efficiency of our proposed WATUNet through the application of a comprehensive range of performance metrics, including the Dice coefficient, accuracy, loss, Dice loss, precision, sensitivity (recall), specificity, F1, and recall. The evaluation was carried out on both the BUSI dataset and our VSI dataset. Furthermore, we conducted a comparative analysis of our network with other contemporary state-of-the-art architectures. This section presents evaluation metrics by utilizing equations (A-1) to (A-7) in Appendix A and reports the quantitative and qualitative results of each component of the framework. The proposed model is compared with other state-of-the-art models by using the same UNet backbone as our model for breast segmentation. The proposed model was developed using Python 3.7 with Keres API and was trained and tested using the NVIDIA RTX A3000 GPU and the NVIDIA Tesla P100 GPU. The model underwent training for 300 epochs using the Adam optimizer (learning rate = 0.001, beta1= 0.9, beta2 = 0.9, epsilon = 0.0000001).

A rigorous analysis of both the quantitative and qualitative outcomes of the proposed framework is conducted. The initial stage presents the results of the training and validation phases. Appendix C describes our model's performance under different input resolutions, as well as an analysis of various activation functions. The impact of applying CLAHE as a preprocessing step is also explored at Appendix C. Finally, after studying these parameters and obtaining the best results, we selected the optimal parameters and fine-tuned the models. We then compared our fine-tuned model to state-of-the-art methods. The same dataset and parameter settings were used, including activation function, loss function, and optimization. This ensured a fair and accurate comparison.

We randomly split the VSI and BUSI datasets into training, validation, and test subsets with the ratio of 80, 10, and 10 percent respectively. The training set is used to train the model by adjusting its parameters based on the input data and the corresponding target values. Upon completion of each epoch, the model is assessed on the validation set to tune the hyperparameters, including but not limited to the learning rate, number of layers, and number of neurons in each layer, to optimize performance. This process is commonly referred to as hyperparameter tuning. Upon conclusion of the training process, the best model was selected based on the validation results and saved as the optimal model. Finally, to perform a performance evaluation, we used a test set that the model had not been exposed to during the training and validation processes.

## 5.1 Quantitative training and validation results on VSI dataset

To select the optimal model for our purposes, we employed a criterion based on the Dice coefficient, which was identified as our primary metric of interest. The Dice coefficient is a robust and reliable metric for evaluating the performance of segmentation models due to its sensitivity to both the TP (True Positive) and FP (False Positive). Accordingly, we saved the weight parameters associated with the best performing model, as determined by the highest value for the Dice coefficient. This process was integral to ensure the selection of the best performing model for our specific objectives.

In Figure 6, we show the results of the training process for the proposed model. The image resolution was set to 128 × 128 with batch size of 32. Convergence of the training and validation accuracy, loss, Dice loss, Dice coefficient, sensitivity (recall), specificity, F1 score, and precision during different epochs indicates that the training process had performed well with no overfitting and/or underfitting of the validation set.

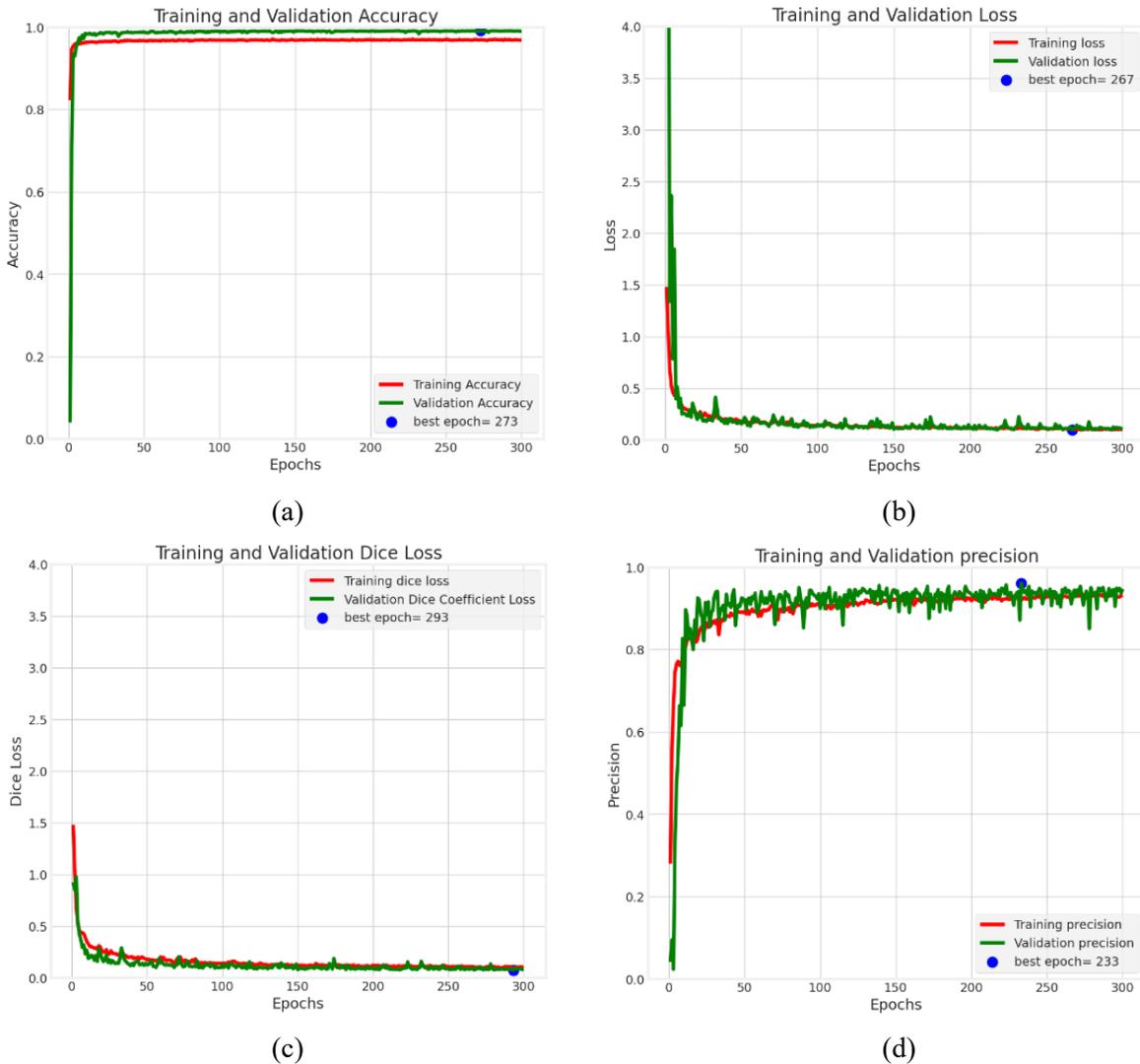

(a) (b)

(c) (d)

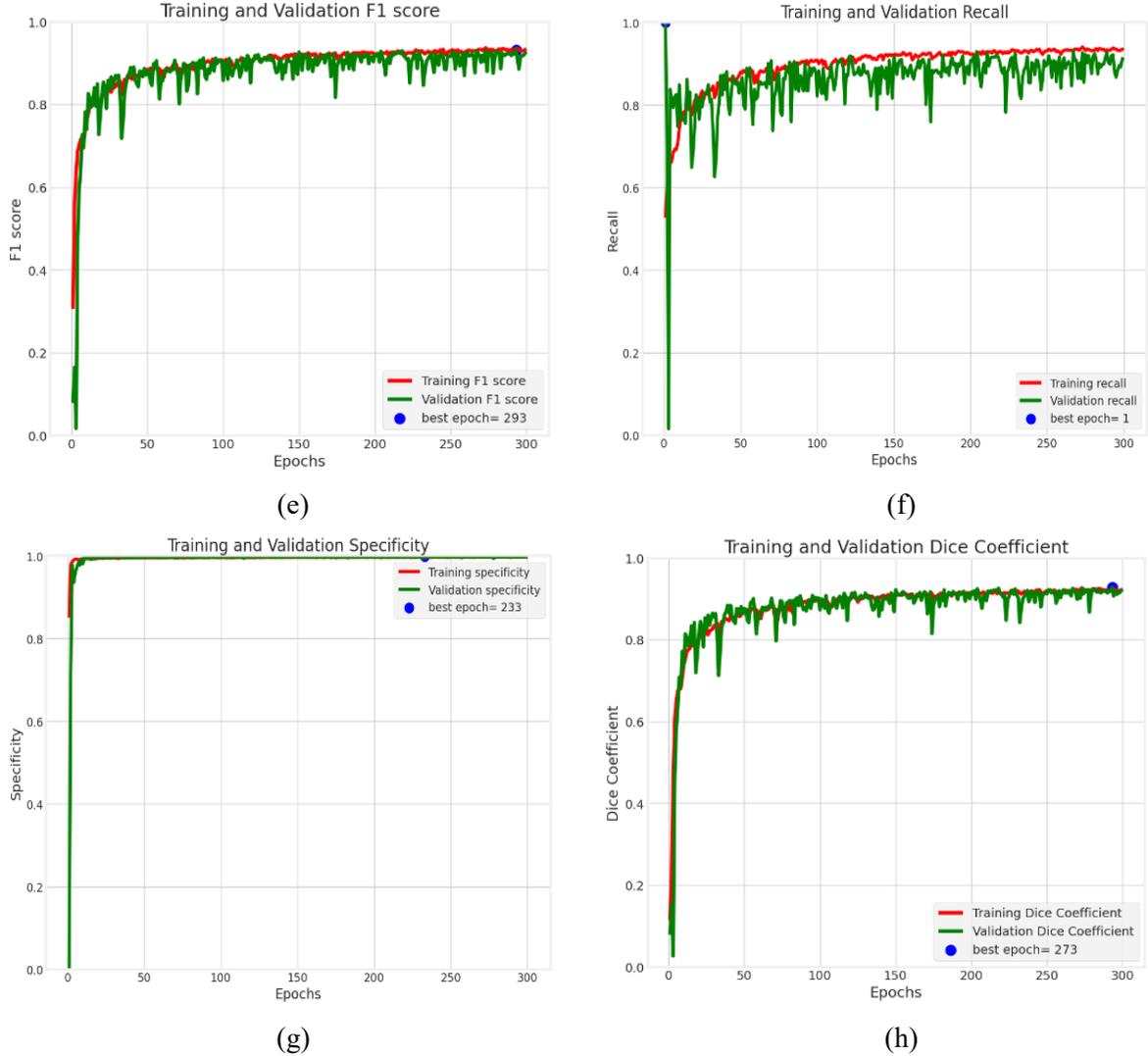

(e)                                                     (f)

(g)                                                    (h)

*Figure 6 WATUNet performance indicators for training and validation. The plots (a-h) represent the performance metrics during training and validation on the VSI dataset.*

## 5.2 Comparative analysis of neural network models on VSI dataset

Table 1 presents the performance metrics obtained by five different neural network models, the backbone of these models are based on previous works namely; UNet (28), Attention UNet (34), Sharp UNet (33), Sharp Attention UNet (35) and the proposed WATUNet model on the test set of the VSI dataset. While the primary network architectures for the four former models have been previously proposed in the literature, the same backbone architecture and hyperparameters were employed for all four models when reporting the results in these tables. The following parameters were identically employed in the training process for all models: image input size of (128 × 128), augmentation technique, optimization technique, batch size, loss function, and learning rate. Each model underwent 300 epochs of training, after which the model exhibiting the best performance on the validation set was selected and saved for comparison on the test set.

The outcome of these comparisons revealed that the WATUNet architecture demonstrated superior performance relative to the other models across all validation parameters.

*Table 1 Comparison of the performance of the proposed WATUNet model with existing ML models using the same settings and parameters - VSI dataset*

| Model | Accuracy | Loss | Dice loss | Precision | F1 | Sensitivity (Recall) | Specificity | Dice coefficient |
|---|---|---|---|---|---|---|---|---|
| UNet | 0.982 | 0.203 | 0.129 | 0.909 | 0.878 | 0.851 | 0.993 | 0.871 |
| Attention UNet | 0.969 | 0.095 | 0.074 | 0.925 | 0.923 | 0.920 | 0.982 | 0.925 |
| Sharpening UNet | 0.984 | 0.166 | 0.106 | 0.924 | 0.900 | 0.878 | 0.994 | 0.894 |
| Sharp Attention UNet (35) | 0.991 | 0.093 | 0.070 | 0.928 | 0.936 | 0.928 | 0.993 | 0.931 |
| UNet++ (29) | 0.988 | 0.195 | 0.115 | 0.905 | 0.896 | 0.888 | 0.995 | 0.884 |
| UNet3+ (31) | 0.988 | 0.178 | 0.109 | 0.935 | 0.897 | 0.863 | 0.997 | 0.890 |
| **Proposed WATUNet** | **0.991** | **0.091** | **0.068** | **0.947** | **0.946** | **0.944** | **0.997** | **0.942** |

The results obtained from the WATUNet algorithm demonstrate an improvement in the sensitivity of lesion area mask extraction for the VSI dataset, with a notable increase of 1.6% compared to the second-best model. The Dice coefficient is also shown to have improved by 1.1% when compared to the second-best model (35). These findings suggest that WATUNet may be a promising solution for improving the accuracy, robustness, and computational efficiency of lesion area segmentation in ultrasound imaging.

We also utilized McNemar's test, which is non-parametric, to compare the relative performance of each segmentation model that we developed. The purpose of the McNemar's test was to assess the statistical significance of differences observed between the models' segmentation results, more details are explained in appendix B. The test was conducted on our VSI test set including 381 images (10%) with dimensions of 128×128 pixels. We compared the dichotomous segmentation results of each pair of models to calculate the number of discordant entries for each pixel in the images. These discordant values were then evaluated using the chi-squared distribution with 1 degree of freedom, which yielded a corresponding p-value.

In hypothesis testing, a p-value below 0.05 is considered significant, allowing us to reject the null hypothesis and provide evidence for the alternative hypothesis. However, in the context of conducting multiple statistical tests simultaneously, e.g., comparing multiple model pairs, the probability of false positives increases. To mitigate this, several correction methods presented, such as the Bonferroni correction which adjusts the alpha value to uphold an appropriate level of significance (53). However, Bonferroni correction can result in a higher rate of Type II errors (false negatives). While the false discovery rate (FDR) correction (54) provides a more balanced trade-off between controlling false positives and false negatives, which can lead to more power to detect significant results.

We employed the Benjamini-Hochberg procedure to obtain adjusted p-values (q-values) by considering the rank of each p-value and the total number of comparisons. We identified statistically significant model pairs by comparing q-values to the specified FDR level. Pairs with q-values ≤ 0.05 were considered significant, indicating meaningful performance differences, while effectively controlling false positives and false negatives for multiple comparisons. Table 2 presents the results of comparing pairs of models after applying the FDR correction to each of the individual models. We calculated the p-value and q-value for each pair, in this Table, *m* is the number of comparisons which is equal to 10. Our proposed model, which combines attention gates and wavelet gates to a UNet framework, exhibited a notable performance improvement over other model.

*Table 2 Statistical significance of model comparisons with Benjamin-Hochberg procedure, where i and m are the rank, and number of comparisons respectively, used to adjust for multiple comparisons.*

| Comparison between Models | p-value | Rank | adjusted p-value = ($i/m$) × alpha | Significant |
|---|---|---|---|---|
| **(WATUNet, Sharpening UNet)** | 7.2e-6 | 1 | **0.005** | Yes |
| **(WATUNet, Sharp Attention UNet (35))** | 7.8e-5 | 2 | **0.010** | Yes |
| (Sharping UNet, UNet) | 2e-5 | 3 | **0.015** | Yes |
| **(WATUNet, Attention UNet)** | 1e-5 | 4 | **0.020** | Yes |
| (Sharp Attention UNet, UNet) | 7e-4 | 5 | **0.025** | Yes |
| (Attention UNet, UNet) | 8.7e-3 | 6 | **0.030** | Yes |
| **(WATUNet, UNet)** | 3.3e-2 | 7 | **0.035** | Yes |
| (Sharpening UNet, Attention UNet) | 1.8e-1 | 8 | 0.040 | No |
| (Sharp Attention UNet, Sharpening UNet) | 2.5e-1 | 9 | 0.045 | No |
| (Sharp Attention UNet, Attention UNet) | 4.2e-1 | 10 | 0.050 | No |

## 5.3 Visual comparison: segmentation model output vs. ground truth data on VSI and BUSI datasets

Figures 7 and 8 illustrate the comparison between the segmentation model output (WATUNet), and the ground truth data (outlined by radiologists) on the VSI and BUSI datasets, respectively. This visual representation provides valuable insight into the efficacy and accuracy of the proposed model's performance. The processed mask is derived by binarizing the predicted mask using a threshold of 0.4. This specific threshold value was determined through the analysis of the receiver operating characteristic (ROC) curve. Figure 9 shows the 3D rendering of a cine sweep and compares the ground truth and the predicted results. The WATUNet proposed segmentation model applied to VSI sweep and the tumor area is extracted, followed by multiplication of the mask with the original image. Then, contouring was applied to each slide. The 3D rendering was achieved using the MATLAB Volume Viewer application. Top views are presented. T at the top-left is an abbreviation for 'Transducer'.

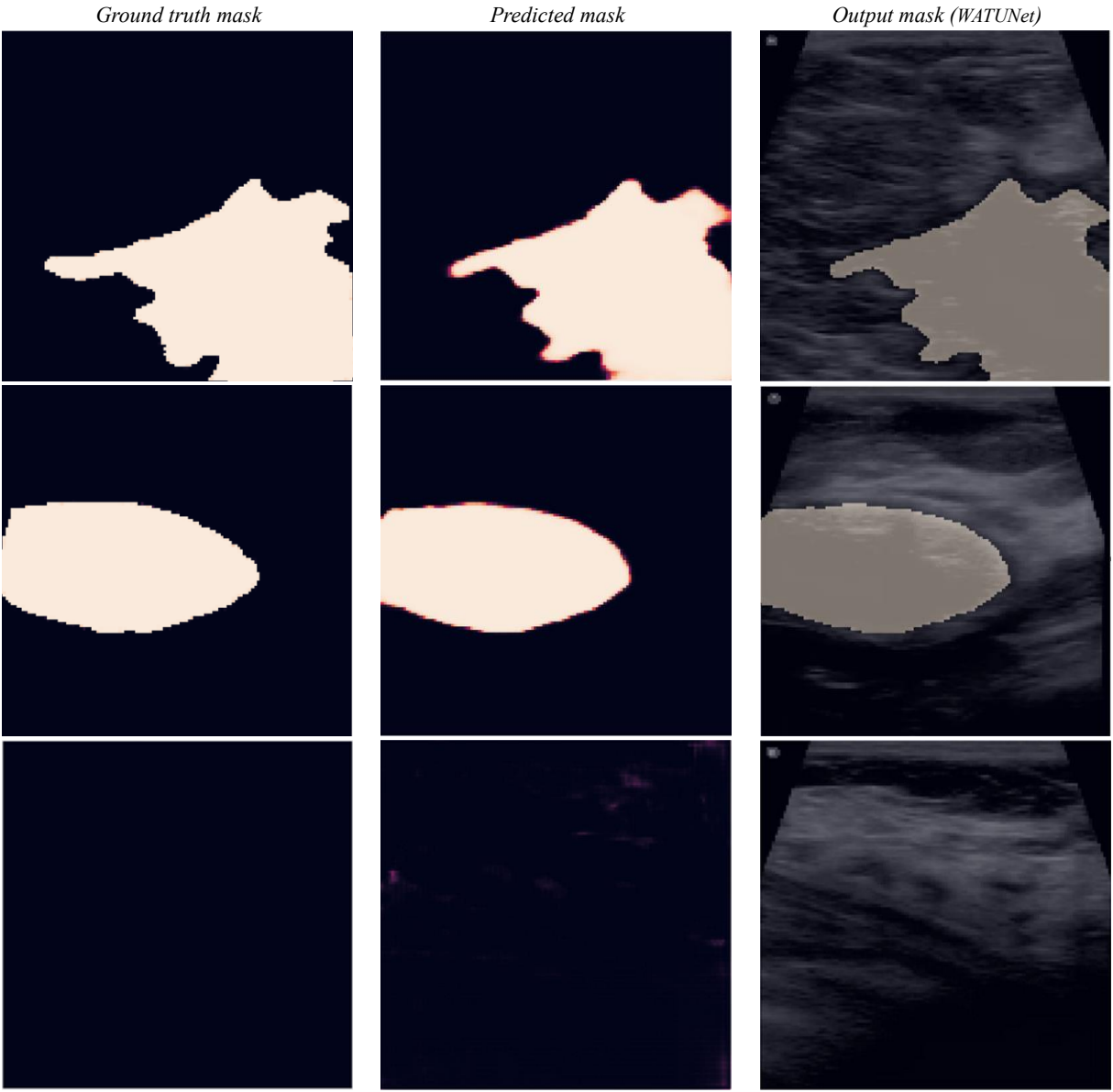

*Figure 7 Some VSI dataset examples with their corresponding ground truth, predicted mask, and WATUNet output mask overlaid on the original images.*
*top row: malignant mass, middle row: benign mass, bottom row: no mass found in the frame.*

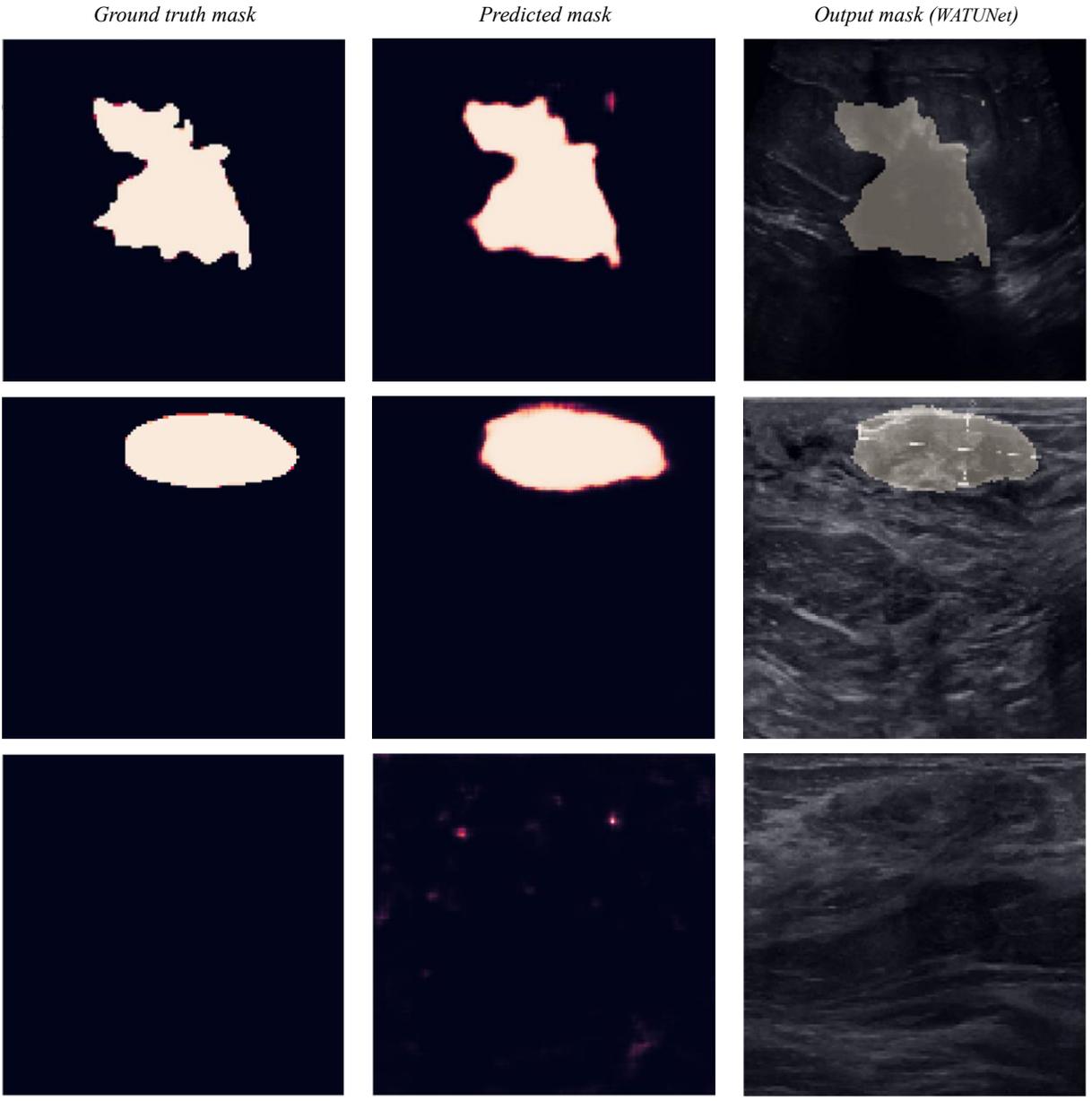

*Figure 8 Some BUSI dataset examples with their corresponding ground truth, predicted mask, and WATUNet output mask overlaid on the original images.*

*top row: malignant mass, middle row: benign mass, bottom row: no mass found in the frame.*

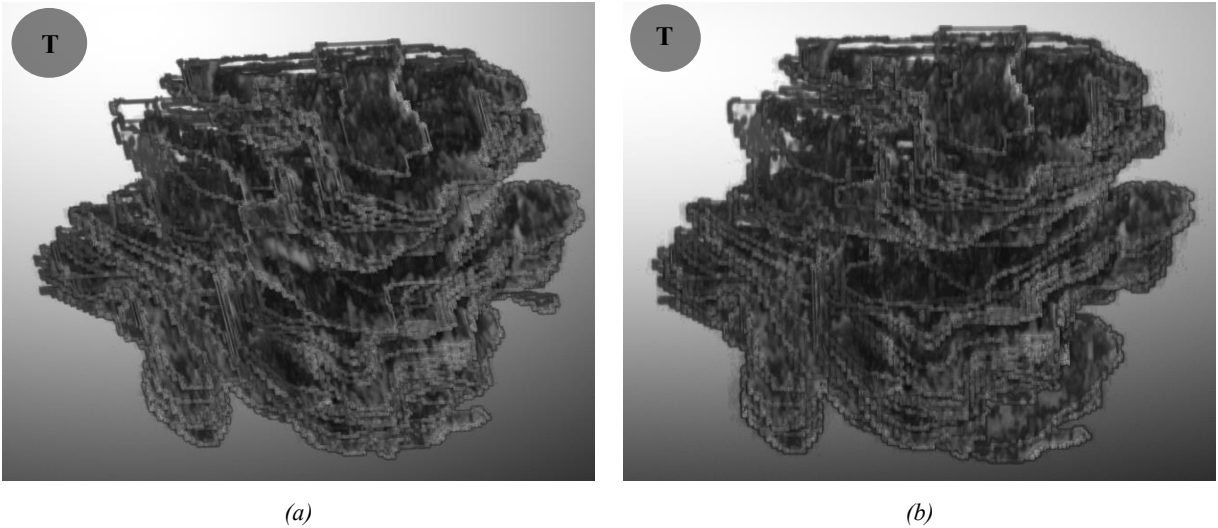

*(a)* *(b)*

*Figure 9 3D rendering highlights tumor area combined with original image, enhanced by contours. (a) Ground truth mask - top view, (b) Predicted mask - top view. (T on the side corner is abbreviation for 'Transducer')*

## 6. Discussion

In this work we proposed a lesion segmentation model, named WATUNet, optimized for volume ultrasound sweeps across breast lesions. It represents an improvement over existing models for the segmentation of ultrasound images by focusing on the skip connections between the encoder and decoder in the UNet architecture, as well as on the UNet backbone, activation functions, regularization, and loss function. This model overcomes many of the limitations of its predecessors. The incorporation of WGs and AGs further enhances the model's ability to capture multi-scale features in both spatial and frequency domains, and to selectively attend to relevant regions in the input image, resulting in improved performance.

We evaluated our model by using two datasets categorized into three classes, including the publicly available "Breast Ultrasound Images" (BUSI) dataset of 780 images and our VSI dataset of 3818 images. The experimental results showed that the proposed WATUNet model outperformed state-of-the-art models in terms of accuracy, loss, specificity, recall, F1, precision, Dice coefficient, and visual representation. These findings indicate that the model holds considerable promise for clinical applications, as it has the potential to improve the accuracy of breast cancer segmentation and reduce the cost associated with traditional methods of tumor detection.

This study's innovation lies in the combination of the VSI scanning approach for breast imaging with deep learning algorithms to achieve highly accurate segmentation of breast lesions. This is particularly relevant to health care in remote areas where access to medical experts may be limited. Our results build on the advantages of using VSI, where the approach's simplicity allows relatively inexperienced operators to obtain high-quality ultrasound images quickly. This contrasts with traditional ultrasound, which requires highly skilled sonographers who may take months or even years to train. By overcoming the need for highly trained sonographers, VSI has the potential to significantly improve access to medical imaging for breast pathology, especially in regions where access to healthcare resources is limited.

By leveraging advanced machine learning techniques, this study offers the potential to improve early detection and segmentation of breast lesions, ultimately leading to improved patient outcomes. Despite the encouraging results, this is a proof-of-concept study. Limitations of this study include the modest size of the two image sets employed, and the two commercial scanners used to obtain these images. Expansions of these need to be addressed in future studies, along with the extension of this work to a final dichotomous diagnosis of the lesion as probably benign or malignant, the next necessary step in deciding the path of patient care. We also plan to further explore wavelet attention-based models and their potential for other applications beyond breast cancer diagnosis. Finally, the integration of VSI and WATUNet would potentially enable an increase in imaging access by allowing for rapid and automatic diagnosis of breast lumps to areas where a radiologist or a skilled sonographer are not present.

**Conflicts of interest**

There is no conflict of interest.

**Acknowledgments**

This work was supported by National Institutes of Health Grants R21EB025290, and R21AG070331.

# Appendix A:

## A.1 BCE_Dice loss

The BCE loss is typically employed in binary classification problems, such as image segmentation, where each pixel is classified as foreground or background. This loss function quantifies the discrepancy between predicted probabilities and true binary labels (as shown in eq. (A-1)). It plays a crucial role in evaluating how well the neural network's predictions align with the actual binary segmentation of the images. The Dice loss, on the other hand, measures the overlap between the predicted segmentation mask and the ground truth mask. It is calculated in eq. (A-3), where the Dice coefficient measures the similarity between the predicted and true mask. The Dice coefficient (eq. (A-4)) quantifies the similarity between the predicted and true masks based on their alignment with each other.

$$BCE_{\text{Dice}} \text{loss} = BCE \text{ loss} + \text{Dice loss} \quad (A\text{-}1)$$

$$BCE \text{ loss} = -\frac{1}{N}\sum_{i=0}^{N}[y_i * \log(p_i) + (1 - y_i) * \log(1 - p_i)] \quad (A\text{-}2)$$

where $y$ is the ground truth label (either 0 or 1), $p$ is the predicted probability of the positive class, and log is a natural logarithm function.

$$\text{Dice loss} = 1 - \text{Dice coefficient} \quad (A\text{-}3)$$

$$\text{Dice coefficient} = \frac{(2 * \text{intersection})}{(\text{size of prediction} + \text{size of ground truth} + \text{epsilon})} \quad (A\text{-}4)$$

The combination of these two loss functions results in generating segmentation masks that are not only precise but also exhibit a significant overlap with the true masks. This approach can enhance the model's performance on segmentation tasks, particularly when faced with sophisticated or unclear boundaries between foreground and background classes. The intersection in the equation represents the number of pixels where both the predicted and ground truth masks have a value of 1. The union represents the number of pixels where either the predicted or ground truth mask has a value of 1. To prevent division by zero, epsilon is a small value (e.g., 1e-5) used in the calculations. When using the $BCE$ loss, training can handle imbalanced foreground and background classes. Since the segmented area of the lesions is smaller than the background, $BCE$ loss is an appropriate approach to address this issue. On the other hand, Dice loss is more sensitive to foreground, emphasizing segmentation accuracy. It is reasonable to take advantage of this sensitivity for appropriate segmentation. Therefore, the $BCE_{\text{Dice}}$ loss penalizes false positives and false negatives through the $BCE$ loss term, while also encouraging overlap between the predicted and ground truth masks through the Dice loss term. The final $BCE_{\text{Dice}}$ loss is the sum of these two terms.

**A.2 Dice coefficient**

The Dice coefficient is a measure of the similarity between two sets. It is defined by eq. (A-4).

**A.3 Accuracy**

Accuracy is a measure of the overall performance of a binary classification model and is calculated as the ratio of correctly predicted pixels to the total number of pixels:

$$\text{Accuracy} = \frac{(TP + TN)}{(TP + FP + TN + FN)} \tag{A-5}$$

where true positive (TP) is the number of pixels that are correctly classified as positive, true negative (TN) is the number of samples that are correctly classified as negative, false positive (FP) is the number of samples that are incorrectly classified as positive, and false negative (FN) is the number of samples that are incorrectly classified as negative.

**A.4 Precision**

Precision is a measure of the proportion of positive predictions that are correct and is calculated as the ratio of TP to the total number of predicted positives of each pixel (TP + FP).

$$\text{Precision} = \frac{TP}{(TP + FP)} \tag{A-6}$$

**A.5 Sensitivity**

Sensitivity, also known as recall or true positive rate (TPR), is a measure of the proportion of actual positive pixels that are correctly identified as positive by the model and is calculated as the ratio of TP to the total number of actual positives of each pixel (TP + FN).

$$\text{Sensitivity} = \frac{TP}{(TP + FN)} \tag{A-7}$$

**A.6 Specificity**

Specificity is a measure of the proportion of actual negative pixels that are correctly identified as negative by the model and is calculated as the ratio of TN to the total number of actual negatives of each pixel (TN + FP).

$$\text{Specificity} = \frac{TN}{(TN + FP)} \tag{A-8}$$

## A.7 F1 Score

The F1 score is a measure of the harmonic mean of precision and sensitivity and provides a balanced evaluation of the model performance.

$$\text{F1 Score} = \frac{2 \times (\text{Precision} \times \text{Sensitivity})}{(\text{Precision} + \text{Sensitivity})} \tag{A-9}$$

## A.8 Jaccard index

The Jaccard Index, also known as intersection over union (IoU), is a measure used to quantify the similarity between two sets.

$$\text{Jaccard index} = \frac{\text{size of intersection}}{\text{size of union}} \tag{A\_10}$$

# Appendix B:

This section provides an explanation of various aspects of the proposed framework, including details regarding the dataset split for evaluating the model and specific implementation details.

## B.1 Quantitative analysis of the proposed WATUNet model for different input image resolutions

The segmentation task on VSI and BUSI images was performed with 3 different network input sizes of 32 × 32, 64 × 64, and 128 × 128. However, attempts to increase image resolutions employing a batch size of 32 were precluded, owing to the excessive memory overhead experienced by the GPU. During the experiment, the number of training epochs was set to 300 and the batch size was set to 32. Table C-1 displays the outcomes of the experiment, exhibiting the performance metrics computed utilizing equations (9) and (A-1) through (A-5). Table C-1 indicates that the input size of 128 × 128 yielded the highest values.

*Table B-1 Quantitative Evaluation of VSI dataset of different input sizes*

| Input resolution | Batch_size | Accuracy | Loss | Dice loss | Precision | F1 | Sensitivity | Specificity | Dice coefficient |
|---|---|---|---|---|---|---|---|---|---|
| 32 × 32 | 32 | 0.989 | 0.130 | 0.094 | 0.877 | 0.872 | 0.868 | 0.996 | 0.906 |
| 64 × 64 | 32 | 0.984 | 0.114 | 0.080 | 0.932 | 0.931 | 0.931 | 0.994 | 0.919 |
| 128 × 128 | 32 | **0.991** | **0.091** | **0.068** | **0.947** | **0.946** | **0.944** | **0.997** | **0.942** |

## B.2 Quantitative analysis of the proposed WATUNet model for different activation functions

In Table B-2, the quantitative analysis conducted on the WATUNet model for different activation functions (i.e., ReLu, LRelu, swish, and mish) revealed that swish outperformed the other functions. This finding suggests that the swish activation function is better suited for the WATUNet model compared to the other evaluated activation functions.

*Table B-2 Quantitative evaluation of VSI dataset with different activation functions*

| Activation Function | Accuracy | Loss | Dice loss | Precision | F1 | Sensitivity | Specificity | Dice coefficient |
|---|---|---|---|---|---|---|---|---|
| ReLu | 0.980 | 0.140 | 0.097 | 0.918 | 0.915 | 0.914 | 0.993 | 0.903 |
| LRelu (alpha=0.1) | 0.982 | 0.120 | 0.084 | 0.928 | 0.927 | 0.927 | 0.993 | 0.916 |
| **Swish** | **0.991** | **0.091** | **0.068** | **0.947** | **0.946** | **0.944** | **0.997** | **0.942** |
| **Mish** | 0.968 | 0.099 | 0.075 | 0.934 | 0.935 | 0.936 | 0.997 | 0.925 |

## B.3 Quantitative analysis of the proposed WATUNet model before and after data augmentation on both VSI and BUSI dataset

In Table B-3, we conducted a comprehensive quantitative analysis of the proposed WATUNet model on both the VSI and BUSI datasets, comparing its performance before and after the implementation of data augmentation techniques. The results revealed improvements in the model's performance following augmentation. Without augmentation, the model is more prone to memorizing the training data rather than learning meaningful patterns. As a result, there can be a significant discrepancy between the training and validation results. The augmented dataset allowed for a more robust and comprehensive training of the model. These findings highlight the effectiveness of data augmentation in improving the performance and generalizability of the WATUNet model. The results show that the baseline VSI dataset displays a higher degree of similarity to the augmented VSI dataset, primarily due to its larger size when compared to the BUSI dataset.

*Table B-3 Quantitative evaluation of applying data augmentation to VSI and BUSI datasets*

| Dataset | Accuracy | Loss | Dice loss | Precision | F1 | Sensitivity | Specificity | Dice coefficient |
|---|---|---|---|---|---|---|---|---|
| Augmented VSI | **0.991** | **0.091** | **0.068** | **0.947** | **0.946** | **0.944** | **0.997** | **0.942** |
| Baseline VSI | 0.988 | 0.155 | 0.109 | 0.924 | 0.895 | 0.871 | 0.997 | 0.891 |
| Augmented BUSI | **0.983** | **0.114** | **0.081** | **0.932** | **0.931** | **0.931** | **0.994** | **0.919** |
| Baseline BUSI | 0.870 | 1.884 | 0.556 | 0.325 | 0.448 | 0.752 | 0.881 | 0.444 |

## B.4 Quantitative analysis of the proposed WATUNet model before and after CLAHE preprocessing on both VSI and BUSI dataset

To analyze the effectiveness of CLAHE, the proposed WATUNet model was evaluated on the test sets utilizing an image size of 128 × 128, a batch size of 32, and employing swish as the activation function. To ascertain the model's efficacy, an assessment was conducted on both the original test sets and CLAHE-enhanced test sets for both the VSI and BUSI datasets. The results in Table B-4 indicate that the application of CLAHE had a positive impact on the outcome, resulting in significant improvements.

*Table B-4 Quantitative evaluation of applying CLAHE to VSI and BUSI dataset*

| Dataset | Accuracy | Loss | Dice loss | Precision | F1 | Sensitivity | Specificity | Dice coefficient |
|---|---|---|---|---|---|---|---|---|

| | | | | | | | |
|---|---|---|---|---|---|---|---|
| **Original VSI** | 0.982 | 0.112 | 0.077 | 0.933 | 0.933 | 0.934 | 0.994 | 0.923 |
| **Original BUSI** | 0.979 | 0.155 | 0.107 | 0.914 | 0.907 | 0.903 | 0.993 | 0.893 |
| **Enhanced VSI** | **0.991** | **0.091** | **0.068** | **0.947** | **0.946** | **0.944** | **0.997** | **0.942** |
| **Enhanced BUSI** | **0.983** | **0.114** | **0.081** | **0.932** | **0.931** | **0.931** | **0.994** | **0.919** |

## B.5 Comparative analysis of neural network models: performance evaluation and superiority of the proposed architecture on the BUSI dataset

In table B-5, we compared the performance of the proposed WATUNet algorithm to the other state of the art UNet-based segmentation methods. WATUNet exhibited improvements on the BUSI dataset, with increases of 0.4% and 1.1% observed for sensitivity and Dice coefficient, respectively.

*Table B-5 Comparison of the performance of WATUNet with some existing ML models with all the same settings and parameters on the BUSI dataset*

| Model | Accuracy | Loss | Dice loss | Precision | F1 | Sensitivity | Specificity | Dice coefficient |
|---|---|---|---|---|---|---|---|---|
| **UNet** | 0.982 | 0.231 | 0.139 | 0.911 | 0.867 | 0.834 | 0.994 | 0.861 |
| **Attention UNet** | 0.984 | 0.166 | 0.106 | 0.924 | 0.900 | 0.878 | 0.994 | 0.894 |
| **Sharpening UNet** | **0.984** | 0.186 | 0.123 | 0.885 | 0.886 | 0.890 | 0.992 | 0.877 |
| **Sharp Attention UNet(35)** | 0.979 | 0.109 | **0.072** | 0.941 | 0.941 | 0.943 | 0.994 | 0.928 |
| **WATUNet model** | 0.983 | **0.104** | 0.078 | **0.942** | **0.945** | **0.947** | 0.994 | **0.939** |